\documentclass[aps,prd,twocolumn,groupedaddress,showpacs]{revtex4-1}
\usepackage{amsmath,amssymb,graphicx}
\usepackage{epstopdf}
\newcommand{\sbar}[2]{\overline{s}^{\text{\tiny #1#2}}}
\newcommand{\sab}{\sbar{A}{B}}
\newcommand{\sjk}{\sbar{J}{K}}
\newcommand{\skl}{\sbar{K}{L}}
\newcommand{\szk}{\sbar{Z}{K}}
\newcommand{\stt}{\sbar{T}{T}}
\newcommand{\sxx}{\sbar{X}{X}}
\newcommand{\sxy}{\sbar{X}{Y}}
\newcommand{\sxz}{\sbar{X}{Z}}
\newcommand{\syy}{\sbar{Y}{Y}}
\newcommand{\syz}{\sbar{Y}{Z}}
\newcommand{\szz}{\sbar{Z}{Z}}
\newcommand{\sigj}{\hat{\sigma}^{\text{\tiny J}}}
\newcommand{\sigk}{\hat{\sigma}^{\text{\tiny K}}}
\newcommand{\sigl}{\hat{\sigma}^{\text{\tiny L}}}
\newcommand{\nj}{\hat{n}^{\text{\tiny J}}}
\newcommand{\spq}{\overline{s}_{\text{\tiny PQ}}}
\newcommand{\spk}{\overline{s}_{\text{\tiny P}k}}
\newcommand{\sqk}{\overline{s}_{\text{\tiny Q}k}}
\newcommand{\ags}{\alpha_{\text{\tiny GS}}}
\newcommand{\dgs}{\delta_{\text{\tiny GS}}}
\newcommand{\egs}{\hat{e}_{\text{\tiny GS}}}
\newcommand{\ens}{\hat{e}_{\text{\tiny NS}}}
\newcommand{\ewe}{\hat{e}_{\text{\tiny WE}}}
\newcommand{\omgs}{\omega_{\text{\tiny GS}}}
\newcommand{\omns}{\omega_{\text{\tiny NS}}}
\newcommand{\omwe}{\omega_{\text{\tiny WE}}}
\newcommand{\omt}{\omega_{\text{\tiny T}}}
\newcommand{\stjk}{\overline{s}_t^{\text{\tiny JK}}}
\newcommand{\stkl}{\overline{s}_t^{\text{\tiny KL}}}
\def\etal {{\it et al.}}

\begin{document}

\title{Limits on violations of Lorentz Symmetry from Gravity Probe B}

\author{Quentin G. Bailey}
\email{baileyq@erau.edu}
\affiliation{Department of Physics, Embry-Riddle Aeronautical University, Prescott, AZ 86301}

\author{Ryan D. Everett}
\author{James M. Overduin}
\email{joverduin@towson.edu}
\affiliation{Department of Physics, Astronomy and Geosciences, Towson University, Towson, MD 21252}

\date{\today}

\begin{abstract}
Generic violations of Lorentz symmetry can be described by an effective field theory framework that contains both general relativity and the standard model of particle physics called the Standard-Model Extension (SME).
We obtain new constraints on the gravitational sector of the SME using recently published final results from Gravity Probe~B.
These include for the first time an upper limit at the $10^{-3}$ level on the time-time component of the new tensor field responsible for inducing local Lorentz violation in the theory, and an independent limit at the $10^{-7}$ level on a combination of components of this tensor field.
\end{abstract}


\maketitle

\section{Introduction}

The two leading approaches to the challenge of unifying the fundamental interactions, string theory and loop quantum gravity, involve extending space to higher dimensions and discretizing spacetime, respectively.
Spontaneous violations of Lorentz symmetry can appear in some versions of the former \cite{KS89}, while the latter can violate Lorentz symmetry explicitly since it entails fixed scales of length or time \cite{GJ99}.  
Experimental tests of Lorentz symmetry have lately gained tremendous traction with the introduction of a comprehensive effective field theory framework for the study, evaluation and comparison of models for Lorentz violation in all sectors of both General Relativity (GR) and the Standard Model of particle physics: the Standard-Model Extension or SME \cite{CK9798}.

The full SME includes all possible Lorentz-violating couplings of background tensor fields to curvature, torsion, and matter fields.
We focus here on the minimal pure-gravity sector of the theory \cite{K04,BK06}, whose action includes, besides the usual Einstein-Hilbert term, a new tensor field $s^{\text{\tiny AB}}$ coupled to the traceless part of the Ricci tensor \cite{Footnote1}.
This field induces violations of local Lorentz invariance, acquiring vacuum expectation values $\sab$ which are assumed constant in asymptotically inertial Cartesian coordinates \cite{BK06}.
The symmetry-breaking is assumed to be spontaneous; that is, associated with the state of the system rather than the underlying dynamics \cite{G96BK05}.
There are nine independent coefficients in $\sab$ and the most promising ways of constraining them have been detailed in Refs.~\cite{BK06,B09,TB11}.
Seven are of particular interest in this paper because they affect the motion and orientation of a gyroscope in orbit around a central mass: $\stt,\sxx,\sxy,\sxz,\syy,\syz$ and $\szz$.
Strong upper bounds have been placed on five different linear combinations of the six spatial coefficients using atom interferometry and lunar laser ranging \cite{B07,M08,C09,AR11}.
Other limits have been investigated based on short-range gravity tests \cite{BSL10} and solar-system orbital constraints \cite{I12}.
Simulations have also been performed with Doppler tracking of the Cassini spacecraft \cite{H12}.
However, the time-time coefficient $\stt$ has remained unconstrained.

In Ref.\ \cite{BK06} it was shown that Gravity Probe~B (GPB) would be sensitive to combinations of the seven coefficients above via its measurement of the geodetic and frame-dragging effects.
We check this using the recently published GPB final results \cite{E11} and obtain an upper limit on $\stt$.
We also find that the new constraint breaks an algebraic degeneracy among existing experimental limits, enabling us for the first time to extract individual upper bounds on {\em all\/} the SME coefficients in the pure-gravity sector.

Our paper is organized as follows: Sec.~\ref{sec:gpb} provides a brief review of gyroscopic tests of gravitational theories and summarizes the experimental results from GPB.
Our constraints (based on the framework of Ref.~\cite{BK06}) are derived in Sec.~\ref{sec:main}.
In Sec.~\ref{sec:rescaling} we consider the effects of a rescaling of Newton's gravitational constant $G$ in the theory, and Sec.~\ref{sec:orbit} examines the effects of orbital perturbations on a circular orbit.
We discuss additional effects from aberration and light-bending in Sec.\ \ref{sec: add effects}.
Sec.~\ref{sec:discussion} is a discussion.
Following Sec.~V of Ref.~\cite{BK06}, we adopt standard Sun-centered celestial equatorial coordinates and label Cartesian coordinates in this frame with capital Latin letters such that $A,B,C,...$ are spacetime indices while $J,K,L,...$ refer to space only. 
Physical units are assumed except where otherwise noted.

\section{Gyroscopic tests} \label{sec:gpb}

The geodetic effect, first investigated by Willem de~Sitter, Jan Schouten and Adriaan Fokker beginning in 1916, provides the sixth experimental test of GR (after the three ``classical tests,'' radar time delay and pulsar binaries) and the first to involve the spin of the test body \cite{O10}.
It may be thought of as arising from two separate contributions: one due to space curvature and the other a spin-orbit coupling between the spin of the gyroscope and the ``mass current'' of the central mass (which is moving in the rest frame of the orbiting gyroscope).
The geometric (space curvature) effect arises because the gyroscope's spin vector $\vec{S}$, orthogonal to the plane of the motion, no longer lines up with itself after one complete circuit through curved spacetime around the central mass \cite{T88}.
The spin-orbit effect can be regarded as a gravitational analog of Thomas precession in classical electromagnetism \cite{R69,W70}, though this identification (and the splitting between the two factors) is to a certain extent coordinate-dependent \cite{B94} and some authors argue for different interpretations \cite{MTW73,S74}.

Frame-dragging, first studied by Hans Thirring and Josef Lense in 1918, provides the seventh experimental test of GR and the first to involve the spin, not only of the test body, but of the source of the field as well.
It arises due to the {\em spin-spin\/} coupling between these two masses, and is the gravitational analog of the interaction between a magnetic dipole and an external magnetic field, or the hyperfine interaction between electron and nuclear spin in atomic physics.
(The corresponding analog of geodetic precession is the interaction between electron spin and orbital angular momentum associated with atomic fine structure \cite{H01}.)
Also known as the Lense-Thirring effect, frame-dragging plays an important role in astrophysics and cosmology \cite{O10,T88}, but in the field of the Earth it is exceedingly weak, and more than two orders of magnitude weaker than the geodetic effect.

\begin{figure}
\begin{center}
%Use following for color figure (online only):
\includegraphics[width=\columnwidth]{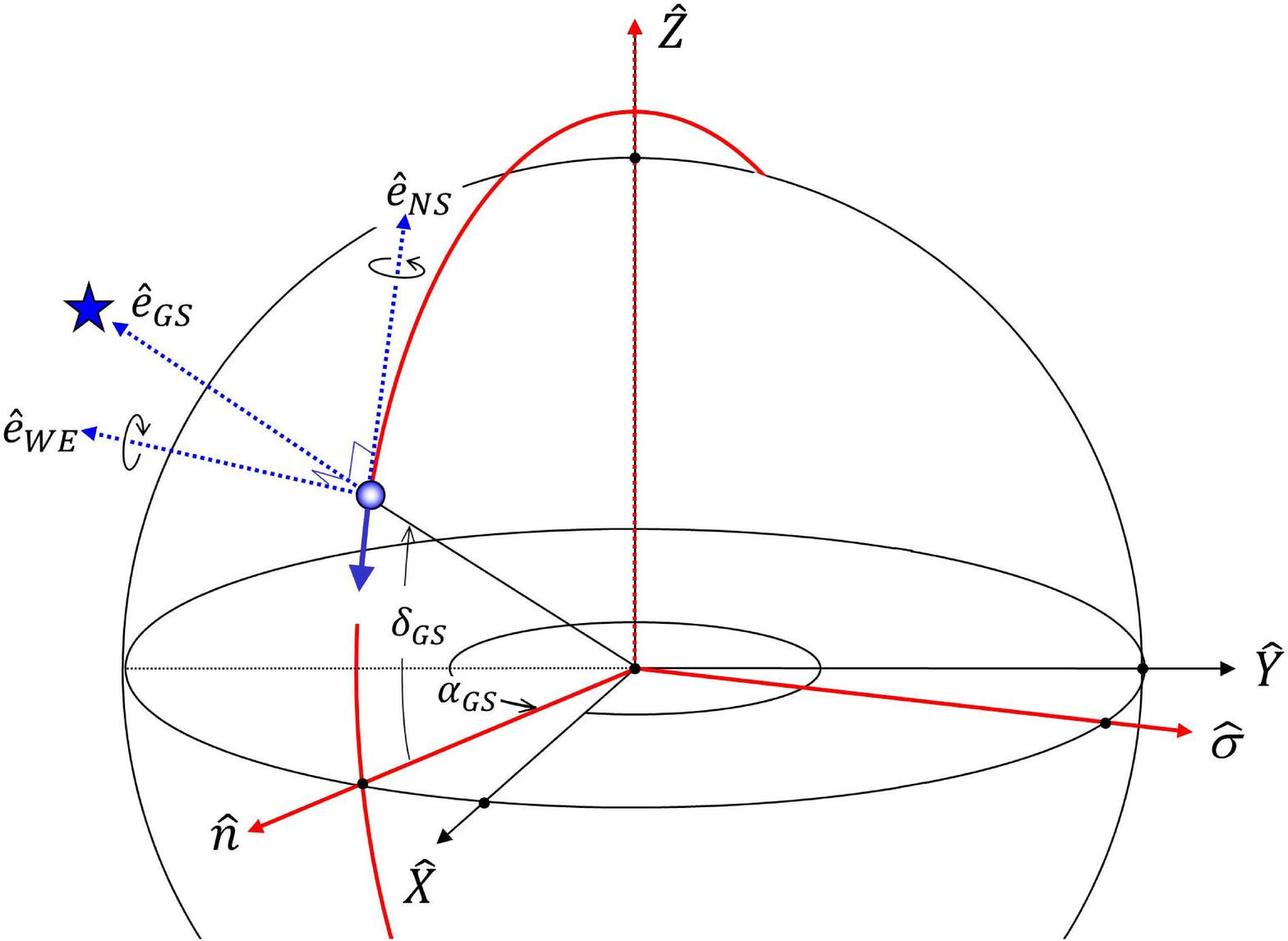}
%Use following for black and white figure (print):
%\includegraphics[width=\columnwidth]{coordinatesCropBW.eps}
\caption{\label{fig:coords}Experimental results are expressed in GPB coordinates $(\egs,\ens,\ewe)$ where $\egs$ points toward the guide star (located in the orbit plane at right ascension $\ags$ and declination $\dgs$), $\ewe$ is an orbit normal pointing along the cross-product of $\egs$ and the unit vector $\hat{Z}$ of the inertial JE2000 frame (aligned with the Earth's rotation axis) and $\ens$ is a tangent to the orbit directed along $\ewe\times\egs$ \cite{E11}.
The theoretical SME predictions are derived in Ref.~\cite{BK06} using inertial $(X,Y,Z)$ coordinates where $\hat{X}$ points toward the vernal equinox and $\hat{Y}=\hat{Z}\times\hat{X}$.
They are subsequently projected onto a hybrid coordinate system ($\hat{\sigma},\hat{Z},\hat{n}$) aligned with the orbit plane, where $\hat{\sigma}=-\ewe$ and $\hat{n}=\hat{\sigma}\times\hat{Z}$.}
\end{center}
\end{figure}

Within GR the geodetic and frame-dragging precession rates of a gyroscope with position $\vec{r}$ and velocity $\vec{v}$ in orbit around a central mass $M$ with moment of inertia $I$ and angular velocity $\vec{\omega}$ are \cite{S60}:
\begin{eqnarray}
\vec{\Omega}_{\text{g,\tiny{GR}}} & = & \left(\frac{3}{2}\frac{GM}{c^2r^3}\right)\vec{r}\times\vec{v} \; , \nonumber \\
\vec{\Omega}_{\text{fd,\tiny{GR}}} & = & \frac{GI}{c^2r^3}\left[\frac{3\vec{r}}{r^2}(\vec{\omega}\cdot\vec{r})-\vec{\omega}\right]
\; .
\label{GRprecessions}
\end{eqnarray}
The combined precession $\vec{\Omega}_{\text{\tiny GR}}=\vec{\Omega}_{\text{g,\tiny{GR}}}+\vec{\Omega}_{\text{fd,\tiny{GR}}}$ causes the unit spin vector $\hat{S}$ of the gyroscope to undergo a {\em relativistic drift\/} given by
\begin{equation}
\vec{R} \equiv \frac{d\hat{S}}{dt}=\vec{\Omega}_{\text{\tiny GR}}\times\hat{S} \; .
\label{GRdrift}
\end{equation}
(In engineering parlance the term ``drift'' connotes an unwanted disturbance, but we use it here to distinguish the desired relativistic signal from unwanted classical disturbances on the gyroscope.)
Averaging over a circular, polar orbit of radius $r_0$ around a spherically symmetric central mass, Eqs.~(\ref{GRprecessions}) simplify to
\begin{equation}
\vec{\Omega}_{\text{g,\tiny{GR}}} = \frac{3(GM)}{2\,c^2r_0^{5/2}}^{3/2} \!\!\!\!\! \hat{\sigma} \;\;\; , \;\;\;
\vec{\Omega}_{\text{fd,\tiny{GR}}} = -\frac{GI\omega}{2\,c^{2}r_0^3} \, \hat{Z} \; ,
\label{GRgeodeticPrecession}
\end{equation}
where $\hat{\sigma}=\hat{r}\times\hat{v}$ is a unit vector normal to the orbit plane and $\hat{Z}=\hat{\omega}$ is the unit vector along the Earth's rotation axis (Fig. \ref{fig:coords}).
With $\hat{S}$ aligned initially along the direction to the guide star (GS), the corresponding relativistic drift rates are, from Eq.~(\ref{GRdrift}):
\begin{eqnarray}
\vec{R}_{\text{g,\tiny{GR}}} & = & -\frac{3(GM)}{2\,c^2r_0^{5/2}}^{3/2} \!\!\!\!\! \ens \; , \nonumber \\
\vec{R}_{\text{fd,\tiny{GR}}} & = & -\frac{GI\omega\cos\dgs}{2\,c^{2}r_0^3} \, \ewe \; ,
\label{GRgeodeticDrift}
\end{eqnarray}
where $\dgs$ is the declination of the guide star and $\ens$ and $\ewe$ are defined in Fig.~\ref{fig:coords}.
The choice of polar orbit orthogonalizes the two effects so that $\vec{R}_{\text{g,\tiny{GR}}}$ points entirely along $\ens$ and $\vec{R}_{\text{fd,\tiny{GR}}}$ points entirely along $\ewe$.
In what follows, it is helpful to keep in mind that a WE component of precession causes the spin vector to drift in the NS direction, and vice versa.
It is also important to note that there is no third component of precession around the GS or guide-star direction, since the gyroscope spin axes were aligned within arcseconds of the guide star in order to maximize sensitivity to the geodetic and frame-dragging effects.
(In practice, there were brief intervals when this condition was not met---as during the post-flight calibration---but the data taken during such periods was unuseable by definition.)
Thus we expect GPB to be able to constrain at most two new linear combinations of SME coefficients.

For GPB with IM~Pegasi as the guide star, $r_0=7018.0$~km \cite{L07} and $\dgs=16.841^{\circ}$, leading to predicted general relativistic drift rates $R_{\text{g,\tiny GR}}=6606.1$~mas/yr (geodetic) and $R_{\text{fd,\tiny GR}}=39.2$~mas/yr (frame-dragging; mas=milliarcsecond)  \cite{E11}.
(The former value differs slightly from that obtained with Eq.~(\ref{GRgeodeticPrecession}), as it takes into account the actual GPB orbit, whose radius and inclination were affected at the 0.1\% level by non-sphericity of the Earth \cite{AS00}.)
The final results of the GPB experiment using all four gyroscopes with 1$\sigma$ uncertainties are $R_{\text{{\tiny NS},obs}}=6601.8\pm18.3$~mas/yr and $R_{\text{{\tiny WE},obs}}=37.2\pm7.2$~mas/yr \cite{E11}.
Thus the NS and WE components of relativistic drift may deviate from the predictions of GR by at most $|\Delta R_{\text{\tiny NS}}| = |R_{\text{g,\tiny GR}}-R_{\text{{\tiny NS},obs}}| <$~22.6~mas/yr and  $|\Delta R_{\text{\tiny WE}}| = |R_{\text{fd,\tiny GR}}-R_{\text{{\tiny WE},obs}}| <$~9.2~mas/yr.

\section{Preliminary constraints} \label{sec:main}

The precession of a gyroscope within the pure-gravity sector of the SME framework has the same form as in standard GR, Eq.~(\ref{GRdrift}), but with additional ``anomalous'' terms containing contributions from the coefficients for Lorentz violation $\sab$ \cite{BK06}:
\begin{eqnarray}
\Delta\Omega^{\,\text{\tiny J}} & = & gv_0\!\left[\tfrac{9}{8}\!\left(\tilde{i}_{(-1/3)}\stt-\tilde{i}_{(-5/3)}\skl\sigk\sigl\right)\sigj\right.\nonumber\\
& & \left.+\tfrac{5}{4}\,\tilde{i}_{(-3/5)}\sjk\sigk\right] \; .
\label{BK156}
\end{eqnarray}
Here $gv_0=\tfrac{2}{3}\Omega_{\text{g,\tiny GR}}$, $\tilde{i}_{(\beta)}\equiv 1+\beta I/Mr_0^2$ and $\sigj=(-\sin\ags,\cos\ags,0)$ where $\ags=343.26^{\circ}$ is the right ascension of the guide star \cite{Footnote2}.
The factor of $\tfrac{2}{3}$ suggests that Lorentz violation affects the geometric, but not the spin-orbit contribution to the geodetic effect.
This is logical, since the violation arises through the coupling of a new field to the curvature tensor.
Detailed investigation of this issue could build on existing studies of the gravitoelectromagnetic limit of the SME \cite{B10,T12}.

The projections of Eq.~(\ref{BK156}) along $(\hat{\sigma},\hat{Z},\hat{n})$ are given by Eqs.~(158-160) of Ref.~\cite{BK06} as
\begin{eqnarray}
\Delta\Omega_{\sigma} & = & \tfrac{2}{3}\Omega_{\text{g,\tiny GR}}\!\left(\tfrac{9}{8}\,\tilde{i}_{(-1/3)}\stt+\tfrac{1}{8}\,\tilde{i}_{(9)}\sjk\sigj\sigk\right), \nonumber \\
\Delta\Omega_z & = & \tfrac{2}{3}\Omega_{\text{g,\tiny GR}}\!\left(\tfrac{5}{4}\,\tilde{i}_{(-3/5)}\szk\sigk\right), \nonumber \\
\Delta\Omega_n & = & \tfrac{2}{3}\Omega_{\text{g,\tiny GR}}\!\left(\tfrac{5}{4}\,\tilde{i}_{(-3/5)}\sjk\nj\sigk\right) \; , \label{BK158-160}
\end{eqnarray}
where $\nj=(\cos\ags,\sin\ags,0)$ \cite{Footnote3}.

Expanding the vector products, collecting terms and simplifying, Eqs.~(\ref{BK158-160}) can be expressed in terms of the individual $\sab$ coefficients as
\begin{equation}
\Delta\vec{\Omega}=\left(\begin{smallmatrix}
\begin{aligned}
& \omt\stt+\omns(\sxx\sin^2{\ags}\\
    &\hspace{7mm}- \sxy\sin{2\ags}+\syy\cos^2{\ags}) \label{BKconstraints}\\ 
& \omwe(\syz\cos{\ags}-\sxz\sin{\ags})\\
& \tfrac{1}{2}\,\omgs(\syy-\sxx)\sin{2\ags}\\
   &\hspace{7mm}+\omgs\sxy\cos{2\ags}\\
\end{aligned}
\end{smallmatrix}\right) \; ,
\end{equation}
where $\omt=\tfrac{3}{4}(1-I/3Mr_0^2)\,\Omega_{\text{g,\tiny GR}}$ = 4503~mas/yr, $\omns=\tfrac{1}{12}(1+9I/Mr_0^2)\,\Omega_{\text{g,\tiny GR}}$ = 1904~mas/yr and $\omwe=\omgs=\tfrac{5}{6}(1-3I/5Mr_0^2)\,\Omega_{\text{g,\tiny GR}}$ = 4603~mas/yr.

To transform from $(\hat{\sigma},\hat{Z},\hat{n})$ coordinates to those used in the GPB data analysis, we reflect across the orbit plane (to carry $\hat{\sigma}$ into $\ewe$) and rotate about $\hat{\sigma}$ by $\dgs$ (to carry $\hat{n}$ into $\egs$).
The resulting components of anomalous precession along the NS, WE and GS axes are
\begin{equation}
\Delta\vec{\Omega}=\left(\begin{smallmatrix}
\begin{aligned}
& -\omt\stt-\omns(\sxx\sin^2\ags\\
   &\hspace{7mm}-\sxy\sin{2\ags}+\syy\cos^2\ags)\\
& \omwe\!\left[\tfrac{1}{2}(\sxx-\syy)\sin{2\ags}\sin{\dgs}\right.\\
   &\hspace{7mm}-\sxy\cos{2\ags}\sin{\dgs}\\
   &\hspace{7mm}-\left.\sxz\sin\ags\cos\dgs\right.\\
   &\hspace{7mm}+\left.\syz\cos\alpha\cos\dgs\right]\\
& \omgs\!\left[\tfrac{1}{2}(\syy-\sxx)\sin{2\ags}\cos{\dgs}\right.\\
   &\hspace{7mm}+\sxy\cos{2\ags}\cos{\dgs}\\
   &\hspace{7mm}-\left.\sxz\sin{\ags}\sin{\dgs}\right.\\
   &\hspace{7mm}+\left.\syz\cos{\ags}\sin{\dgs}\right]\\
\end{aligned}
\end{smallmatrix}\right) \; .
\end{equation}
The corresponding components of anomalous relativistic drift are obtained by taking the cross-product of $\Delta\vec{\Omega}$ with $\hat{S}=\egs$, as in Eq.~(\ref{GRdrift}).
Putting in the numbers, we obtain (in mas/yr):
\begin{eqnarray}
\Delta R_{\text{\tiny NS}} & = &
-4503 \stt - 158 \sxx - 1050 \sxy - 1746 \syy \; , \nonumber \\
\Delta R_{\text{\tiny WE}} & = &
-368 \sxx - 1112 \sxy + 1269 \sxz + 368 \syy\nonumber\\
   & & + 4219 \syz \; , \nonumber \\
\Delta R_{\text{\tiny GS}} & = & 0 \; . \label{numericalLimits}
\end{eqnarray}
In the same units, GPB tells us that $|\Delta R_{\text{\tiny NS}}|<22.6$ and $|\Delta R_{\text{\tiny WE}}|<9.2$.
As discussed above, there is no component of drift around the direction to the guide star, since this is also the direction of the gyroscope spin axes.
The experiment has been designed for optimal measurement of the geodetic and frame-dragging effects.
Of necessity, this entails a loss of sensitivity to any possible third component of precession orthogonal to the other two. 
Thus, in principle we have two constraint equations for seven unknown coefficients $\stt,\sxx,\sxy,\sxz,\syy,\syz,\szz$.

To obtain upper limits on all seven coefficients we turn to the literature for additional  constraints from experiment \cite{AR11}.
At present there are five of these, from combined analysis of atom interferometry and lunar laser ranging \cite{C09}:
\begin{eqnarray}
& & |\sxy| < 2.1 \times 10^{-9} \; , \label{xyConstraint} \\
& & |\sxz| < 4.1 \times 10^{-9} \; , \label{xzConstraint} \\
& & |\syz| < 2.0 \times 10^{-9} \; , \label{yzConstraint} \\
& & |\sxx - \syy| < 2.8 \times 10^{-9} \; , \label{xx-yyConstraint} \\
& & |\sxx + \syy - 2\szz| < 39.8 \times 10^{-9} \; . \label{xx+yy-2zzConstraint}
\end{eqnarray}
Thus it appears that we may have seven equations in seven unknowns.
However, $\Delta R_{\text{\tiny WE}}$ is a linear combination of $\sxx-\syy,\sxy,\sxz$ and $\syz$.
Hence the WE or frame-dragging constraint from GPB is equivalent to a linear combination of Eqs.~(\ref{xyConstraint}-\ref{xx-yyConstraint}), and is moreover superseded by them (since it is about six orders of magnitude weaker).
It is worth noting that the GS component of $\Delta\vec{\Omega}$ consists of another linear combination of the same constraints, so that the GS component of relativistic drift would also not provide an additional constraint in practice, even if it could do so in principle.

We are left with only the NS or geodetic-effect constraint (\ref{numericalLimits}) from GPB, making a total of six experimental constraints on seven unknowns.
The new limit from GPB may be expressed as:
\begin{equation}
|\stt+0.035\sxx +0.39\syy+0.23\sxy| < 5.0\times10^{-3} \; ,
\label{gpbConstraint}
\end{equation}
consistent with pre-GPB estimates of between $5\times10^{-4}$ \cite{BK06} and $\sim10^{-2}$ \cite{O07}.
Although it is quantitatively weaker than existing upper limits from atom interferometry, Eq.~(\ref{gpbConstraint}) is qualitatively important for two reasons: first, because it is the first experimental constraint on the {\em time-time\/} coefficient $\stt$ \cite{AR11}.
Second, it allows us to break the degeneracy between existing constraints.
We need only one additional relationship between the SME coefficients, which comes from the requirement that $\sab$ must be traceless \cite{K04}:
\begin{equation}
\stt - \sxx - \syy - \szz = 0 \; .
\label{tracelessCondition}
\end{equation}
(This condition arises because the Lorentz-violating tensor field $s^{\mu\nu}$ couples to the trace-free Ricci tensor in the action of the theory.
Physically, it reflects the fact that an unobservable overall scaling factor can be removed from the theory.)

Equations~(\ref{xyConstraint}-\ref{tracelessCondition}) together provide us with seven linearly independent equations that we can use to constrain all the SME coefficients {\em individually\/} for the first time.  To see this it is convenient to re-express Eq.~(\ref{gpbConstraint}) in terms of experimentally accessible combinations of the $\sab$ parameters, using the traceless condition~(\ref{tracelessCondition}). When this is done, the GPB constraint reads
\begin{eqnarray}
&& |\stt - 0.15 (\sxx-\syy) + 0.062 (\sxx+\syy-2\szz) \nonumber\\
&& +0.20 \sxy| < 4.4 \times 10^{-3} \; .
\label{gpbConstraint2}
\end{eqnarray}
In combination with the constraints (10-14) from atom interferometry and lunar laser ranging, we then find that $|\stt| < 4.4 \times 10^{-3}$ while $|\sxx|,|\syy|$ and $|\szz|$ are all less than $1.5 \times 10^{-3}$ \cite{O13}.

\section{Rescaling of Newton's gravitational constant} \label{sec:rescaling}

There are two approximations built into the treatment above.
The first involves a rescaling of the effective gravitational constant $G$, which is expected to affect primarily the $\stt$ coefficient since this always appears together with $GM$ in the equations of motion in the SME \cite{BK06}.
To see this, we note that all the anomalous drift rates~(\ref{BK158-160}) in the theory depend on the gravitational mass of the central body through the dimensional factor $\Omega_{\text{g,\tiny GR}}$.
To be fully self-consistent, we must deduce the value of $GM$ from the way that a test body behaves in the vicinity of the central mass.

The equation of motion, at the Newtonian level of approximation, is
\begin{eqnarray}
\frac{d^2r^{\text{\tiny J}}}{dt^2} & = & -\frac{GM}{r^3}\left[(1+\tfrac{3}{2}\stt)r^{\text{\tiny J}}-\sjk r^{\text{\tiny K}}+\tfrac{3}{2}\skl\hat{r}^{\text{\tiny K}}\hat{r}^{\text{\tiny L}}r^{\text{\tiny J}} \right] \nonumber \\
   & & +\frac{GI}{r^5}\left[-\sjk r^{\text{\tiny K}}-\tfrac{5}{2}\skl\hat{r}^{\text{\tiny K}}\hat{r}^{\text{\tiny L}}r^{\text{\tiny J}} \right] \; ,
\label{eom}
\end{eqnarray}
from Eq.~(162) of Ref.~\cite{BK06}.
The Earth's inertia $I$ terms have been added since they are relevant when the orbit is close to the surface.
Here and in the remainder of the paper we set $c=1$ for convenience and express our results in terms of an explicitly traceless version of $\sjk$ defined by
\begin{equation}
\stjk \equiv \sjk - \tfrac{1}{3}\delta^{\text{\tiny JK}}\stt \; ,
\label{sTraceless}
\end{equation}
where $\delta^{\text{\tiny JK}}$ is the usual Kronecker delta.
(This is not a new physical condition, but merely a mathematically convenient way for us to assess the effects of the rescaling on $\stt$.)
Eq.~(\ref{eom}) then becomes
\begin{eqnarray}
\frac{d^2r^{\text{\tiny J}}}{dt^2} & = & -\frac{GM}{r^3}\left[(1+\tfrac{5}{3}\stt)r^{\text{\tiny J}}-\stjk r^{\text{\tiny K}}+\tfrac{3}{2}\stkl\hat{r}^{\text{\tiny K}}\hat{r}^{\text{\tiny L}}r^{\text{\tiny J}} \right] \nonumber \\
   & & +\frac{GI}{r^5}\left[-\stjk r^{\text{\tiny K}}-\tfrac{5}{2}\stkl\hat{r}^{\text{\tiny K}}\hat{r}^{\text{\tiny L}}r^{\text{\tiny J}} \right]\; .
\label{eomTraceless}
\end{eqnarray}
In writing the equation this way, we have separated the {\it isotropic} (or Keplerian) terms, those that merely scale the spherically symmetric acceleration,
from the {\it anisotropic} terms, that can potentially deform the elliptical shape of the orbit.

From this result (\ref{eomTraceless}) it is clear that the effective gravitational mass of the central body is
\begin{equation}
GM^{\prime} = GM(1+\tfrac{5}{3}\stt) \; .
\label{GMprime}
\end{equation}
Inserting Eq.~(\ref{GMprime}) into Eq.~(\ref{BK156}), and again expressing the results in terms of $\stjk$, we obtain a revised expression for the anomalous precession:
\begin{eqnarray}
\Delta\Omega^{\,\text{\tiny J}} & = & \Omega_{\text{g,\tiny GR}}\!\left[-\tfrac{4}{3}\,\tilde{i}_{(0)}\stt-\tfrac{9}{8}\tilde{i}_{(-5/3)}\stkl\hat{\sigma}^{\text{\tiny K}}\hat{\sigma}^{\text{\tiny L}}\hat{\sigma}^{\text{\tiny J}} \right.\nonumber\\
   & &+\left.\tfrac{5}{4}\,\tilde{i}_{(-3/5)}\stjk\hat{\sigma}^{\text{\tiny K}}\right] \; ,
\label{newBK156}
\end{eqnarray}
where the ``$GM$'' in $\Omega_{\text{g,\tiny GR}}$ now refers to the rescaled $GM^{\prime}$.
This is, however, an effective or measured quantity, so its {\em value\/} remains the same as before.
The only change to  Eq.~(\ref{BK156}) occurs in the projection along  $\hat{\sigma}$.
The first of Eqs.~(\ref{BK158-160}) becomes
\begin{equation}
\Delta\Omega_{\sigma} = \tfrac{2}{3}\Omega_{\text{g,\tiny GR}}\!\left(-\tfrac{4}{3}\,\stt+\tfrac{1}{8}\,\tilde{i}_{(9)}\stjk\hat{\sigma}^{\text{\tiny J}}\hat{\sigma}^{\text{\tiny K}}\right) \; .
\end{equation}
Expanding, collecting terms and simplifying terms as before, we find that the first component of $\Delta\vec{\Omega}$ in Eq.~(\ref{BKconstraints}) becomes $\omt\stt+\omns[\sxx(\sin^2\ags-\tfrac{1}{3})-\sxy\sin2\ags+\syy(\cos^2\ags-\tfrac{1}{3})-\tfrac{1}{3}\szz]$.
The only numerical change is to the value of $\omt$, which now reads $-5872$~mas/yr.
The rest of the analysis follows the preceding section.
The second (WE) and third (GS) components of Eqs.~(\ref{numericalLimits}) are unchanged, but the first or NS component is revised to
\begin{eqnarray}
\Delta R_{\text{\tiny NS}} & = &
5872 \stt + 477 \sxx - 1050 \sxy - 1111\syy \nonumber \\
& & + 634 \szz \; .
\label{DeltaRNSrevised}
\end{eqnarray}
We then find that the constraint (\ref{gpbConstraint}), 
once again expressed in terms of the experimentally relevant coefficient combinations, 
is modified to
\begin{eqnarray}
& & |\stt+0.14(\sxx-\syy)-0.054(\sxx+\syy-2\szz) \nonumber\\
& & -0.18\sxy| < 3.8\times10^{-3} \; .
\label{gpbConstraintRevised}
\end{eqnarray}
In combination with Eqs.~(\ref{xyConstraint})-(\ref{xx+yy-2zzConstraint}), we then obtain the slightly stronger upper limits listed in Table~\ref{table:results}.
\begin{table}
\caption{\label{table:results} 1$\sigma$ upper limits on the magnitudes of the SME coefficients, taking into account the rescaling of Newton's constant $G$.}
\begin{ruledtabular}
\begin{tabular}{cc}
Coefficient & Upper Limit\\
\hline
$|\stt|$ & $3.8 \times 10^{-3}$\\
$|\sxx|$ & $1.3 \times 10^{-3}$\\
$|\syy|$ & $1.3 \times 10^{-3}$\\
$|\szz|$ & $1.3 \times 10^{-3}$\\
\end{tabular}
\end{ruledtabular}
\end{table}

\section{Orbital effects} \label{sec:orbit}

The second assumption inherent in the treatment above also relates to the motion of a massive test body in orbit around the central mass.
Our discussion to this point has assumed a circular orbit.
This is an excellent approximation in the case of GPB, whose orbit had an eccentricity of $e=0.00134$ \cite{L07}.
For the terms in Eqs.~(\ref{BK156}) that are already proportional to $\sab$ coefficients, the effects of non-circularity will be insignificant.
The Lorentz-violating coefficients may, however, also perturb the leading-order {\em general-relativistic\/} precessions in Eqs.~(\ref{GRgeodeticDrift}), introducing {\em new\/} $\sab$-dependent terms into the relativistic drift equation that might compete with the anomalous drifts already identified.

Since $\stt$ is associated only with the ``unperturbed'' (Keplerian) ellipse in the equation of motion~(\ref{eomTraceless}), we expect that the other coefficients will play a stronger role.
They will act as perturbing accelerations, distorting the shape of the orbit.
But this will also feed back into our upper limits on all the SME coefficients via the relativistic drift equation, Eq.~(\ref{GRdrift}).
To assess the possible importance of this effect, the most straightforward approach is to look for the effect of secular changes in the orbital elements, due to the coefficients $\sab$, on the precession rate.
Any extra anomalous drift that accumulates can be calculated using Eq.~(\ref{GRdrift}).

We will focus on the effects from the geodetic precession, the first of Eqs.~(\ref{GRprecessions}), 
which can be expected to dominate Lorentz-violating orbital corrections arising from the frame-dragging precession.
The geodetic precession rate, time-averaged over one orbit but generalized to the case of an arbitrary ellipse, is
\begin{equation}
\left\langle \vec{\Omega} \right\rangle_{\text{g,\tiny{GR}}} = \frac{3n^3 a^2 {\hat \sigma}}{2\, (1-e^2)^{3/2}} \; , \label{Omega_avg}
\end{equation}
where $a$, $n$, and $e$ are the semi-major axis, mean frequency, and eccentricity of the orbit.
We now consider possible secular precessions of the orbital elements that appear in this expression.
These precessions can be calculated using the equations of motion~(\ref{eomTraceless}) and the standard perturbative method of ``osculating" elements \cite{S53B91}, which allows for time variation of the six orbital elements specifying an orbit.

For small perturbations on a Keplerian ellipse, the equations for the time derivatives of the orbital elements (e.g., $da/dt$, $de/dt$, etc.) can be averaged over one orbit to obtain the leading secular changes in the elements.
We can then expand around the initial values of the orbital elements in a Taylor series, for which it suffices to truncate the series to first order.
Thus we will use, for example,
\begin{equation}
a = a_0 + \left\langle \frac{da}{dt} \right\rangle t + ... \;\;\; , \;\;\;
e = e_0 + \left\langle \frac{de}{dt} \right\rangle t + ... \; ,
\label{series}
\end{equation}  
where $a_0$ and $e_0$ are the initial values of the semi-major axis and eccentricity, 
and the averaged $da/dt$ and $de/dt$ are to be evaluated also with the initial values.

The secular precessions of the orbital elements for the case of point masses were calculated in Ref.~\cite{BK06}.
We include in our results here the inertia $I$ terms in~(\ref{eomTraceless}).
It will be convenient here to refer to a triad $\{\vec P, \vec Q, \vec k\}$ of orthonormal vectors for a generic elliptical orbit that were used and defined in Ref.~\cite{BK06}.
Briefly, $\vec P$ points along the perigee direction, $\vec k$ points normal to the orbit in the direction of the orbital angular momentum (thus $\hat k=\hat \sigma$ for the gyroscope), and $\vec Q$ points in the orbital plane perpendicular to $\vec P$ ($\vec P \times \vec Q = \vec k$).
For the semi-major axis $a$, there is no change when averaged over one orbit:
\begin{equation}
\left\langle \frac{da}{dt} \right\rangle = 0 \; .
\label{adot}
\end{equation} 
The frequency $n$ is related to the semi-major axis by the relation $n^2 a^3=GM^\prime$, which holds even for the ``osculating" ellipse.
Thus the frequency also does not change.
The secular change in the eccentricity to leading order in the initial eccentricity is given by
\begin{equation}
\left\langle \frac{de}{dt} \right\rangle = \tfrac 14 n \spq e_0 \; ,
\label{edot}
\end{equation}
where the error terms are fourth order in $e_0$.
The subscript $PQ$ stands for the projection of the coefficients along $\vec P$ and $\vec Q$ ($\spq=\sjk P^{\text{\tiny J}} Q^{\text{\tiny K}}$).
Since $e_0=0.00134$, the effect of this secular change on Eq.~(\ref{Omega_avg}) is negligible compared to the $\sab$-dependent terms already present in the expression for the anomalistic drift (\ref{BK156}).
Thus it appears that if there are any relevant secular changes in the orbital elements they must be confined to changes in the orbital angular momentum direction $\vec k = \hat \sigma$.

For a general elliptical orbit, the expression for $\vec k$, in terms of the orbital inclination $i$, and the longitude of the node $\Omega$, 
is
\begin{equation}
\vec k = 
\left(\!\!\begin{array}{c}
\phantom{-}\sin i \sin \Omega \\
-\sin i \cos \Omega \\
\cos i
\end{array}\right) \; ,
\label{k}
\end{equation}
written in terms of the underlying $XYZ$ coordinates in Fig.~\ref{fig:coords}.
The time rate of change of $\vec k$ is
\begin{equation}
\frac {d \vec k}{dt}  = 
\left(\!\!\begin{array}{c}
\phantom{-}\cos i \sin \Omega \frac {di}{dt} +\sin i \cos \Omega \frac {d \Omega}{dt} \\
-\cos i \cos \Omega \frac {di}{dt} +\sin i \sin \Omega \frac {d \Omega}{dt} \\
-\sin i \frac {di}{dt}
\end{array}\right) \; .
\label{kdot}
\end{equation}
The remaining secular changes in the orbital elements that are needed describe the changes in the orientation of the ellipse due to the presence of the coefficients $\sab$.
For the inclination and longitude of the node we obtain to lowest order in eccentricity
\begin{eqnarray}
\left\langle \frac{di}{dt} \right\rangle &=& \tfrac{1}{2} n \left(1+\frac{I}{Ma^2}\right) (\spk \cos \omega - \sqk \sin \omega) \; ,\nonumber\\
\left\langle \frac{d\Omega}{dt} \right\rangle &=& \tfrac{1}{2} n \left(1+\frac{I}{Ma^2}\right) \nonumber \\
   & & \times \csc i (\spk \sin \omega + \sqk \cos \omega) \; .
\label{idot}
\end{eqnarray}
Note that the dependence on the perigee angle $\omega$ actually vanishes when the expressions for $\vec P$, $\vec Q$, and $\vec k$ are inserted into the coefficient combinations $\spk$ and $\sqk$.
Furthermore, we will specialize to the GPB orbit for which $i=\pi/2$ and $\Omega=\ags$.

We can now collect the results obtained into an expression for the anomalous precession due to Lorentz-violating orbital effects.
Denoting this extra precession vector $\Delta\vec{\Omega}^\prime$, we obtain in $XYZ$ coordinates
\begin{eqnarray}
\Delta\vec{\Omega}^\prime &=&
\tfrac 34 n^4 a^2 \left(1+\frac{I}{Ma^2}\right) \, t \nonumber\\
&\times & \! \left(\!\!\!\begin{array}{l}
\phantom{-}\sxz \sin \ags \cos \ags  -\syz \cos^2 \ags \\
\phantom{-}\sxz \sin^2 \ags - \syz \sin \ags \cos \ags \\
-(\sxx-\syy) \sin \ags \cos \ags\\
   \hspace{12mm}+\sxy \cos 2\ags
\end{array}\right) \; ,
\label{DeltaOmegap}
\end{eqnarray}
where $t$ is the time elapsed from the start of the gyroscope orbit.
Note that the explicit appearance of coordinate time $t$ is qualitatively different from two types of precession in the standard GR result.  This implies that the precession rate due the coefficients $\sab$ not only changes with but is also amplified by the duration of the orbiting gyroscope experiment.

The result~(\ref{DeltaOmegap}) can be projected into the GPB coordinates described in Sec.~\ref{sec:gpb}.
There are actually only two linearly independent vectors in $\Delta\vec{\Omega}^\prime$, as the result is perpendicular to the $\vec k$ direction.
Expressed in the $(\egs,\ens,\ewe)$ frame, and upon plugging in the values for the GPB orbit, we obtain
\begin{eqnarray}
\Delta\vec{\Omega}^\prime &=&[ 0.362 (\sxx-\syy) +1.095 \sxy \nonumber \\
   & & -1.249 \sxz - 4.152 \syz ] \, t \, \egs \nonumber \\ 
   & & + [ 1.196 (\sxx-\syy) + 3.616 \sxy \nonumber \\
   & & +0.378 \sxz + 1.257 \syz ] \, t \, \ens \; ,
\label{DeltaOmegap2}
\end{eqnarray}
where $t$ is in seconds and $\Delta\vec{\Omega}^\prime$ is in $\rm mas/yr$.

Because of the dependence of Eq.~(\ref{DeltaOmegap2}) on time $t$, the precession rate is not constant, and we must integrate the first order differential equation for the spin vector, Eq.~(\ref{GRdrift}), over the span of one year to find the extra precession or drift induced by these $\sab$-dependent terms.
The results of this integration can be viewed as an extra drift of the gyroscopic spin along the WE directions which we can add to the results in Eq.~(\ref{numericalLimits}).  Specifically we find that the effects of SME coefficients on the spin precession via orbital perturbations
produce the following extra drift (in mas/yr):
\begin{eqnarray}
\Delta R^{\,\prime}_{\text{\tiny WE}} & = &
-1.89 \times 10^7 ( \sxx -\syy ) -5.71 \times 10^7 \sxy  
\nonumber\\ &&- 5.96 \times 10^6 \sxz -1.98 \times 10^7 \syz \; . \label{extradrift}
\end{eqnarray}
There is no extra drift in the NS direction, as expected, since the precession vector $\Delta\vec{\Omega}$ lacked a component in the WE direction.

Adding this drift to $\Delta R_{\text{\tiny WE}}$ in Eq.~(\ref{numericalLimits}), we find an additional term in the WE constraint from GPB.  In fact, it now becomes strong enough to be almost competitive with existing limits from atom interferometry and lunar laser ranging.
The revised GPB constraint from the WE (frame-dragging) direction reads:
\begin{eqnarray}
& & |(\sxx-\syy)+3.0\sxy+0.32\sxz+1.0\syz| \nonumber \\
   & & < 4.9\times10^{-7} \; .
\label{gpbConstraintRevisedAgain}
\end{eqnarray}
Repeating the analysis of Sec.~\ref{sec:main} together with Eqs.~(\ref{xyConstraint})-(\ref{xx+yy-2zzConstraint}), we find that the frame-dragging constraint is still weaker than existing limits, but now by a factor of only 10 (rather than $10^6$).  Our upper limits on the SME coefficients thus remain unchanged from those in Table~\ref{table:results}.
Nevertheless this gain of some four orders of magnitude in sensitivity highlights the potential importance of frame-dragging as a probe of Lorentz violation through the latter's effects on the gyroscope orbit.

\section{Additional effects: aberration and light bending} \label{sec: add effects}

Other effects enter into the total measured drift as well, and it is worthwhile to ask whether these might also lead to further constraints.  Two examples are aberration and relativistic light deflection, both of which are fully modeled and accounted for in the GPB data analysis, assuming the validity of GR.  In fact, an independent cross-check of this assumption was made when the guide star approached within $22.1$ degrees of the Sun and GPB measured a deflection angle of $21 \pm 7$~mas, in agreement with the GR prediction of $21.7$~mas \cite{E11}.  

A covariant derivation of gyroscope precession, that matches the actual experimental technique of referencing the gyroscopic spin to the incoming light from the guide star, was carried out for GR in Ref.\ \cite{W03}.  It is a priori unclear what happens in this derivation for the modified metric of the SME when the $\sab$ terms are included.  In particular, it is not clear that this alternative derivation should completely match the method used in reference \cite{BK06}, where the spin four-vector $\bf S$ was projected along a comoving but not co-rotating set of spatial vectors ${\bf e}_j$ attached to the satellite.  Some preliminary calculations for the SME, paralleling those in Ref.\ \cite{W03}, show that the relativistic precession for the SME that arises in this alternative derivation matches the previously obtained results in equation (\ref{BK156}), after averaging over one gyroscope orbit.  As in the GR case, light deflection terms arise when the gyro spin vector is projected along the tangent vector to the incoming starlight.  However, the size of the light deflection terms that involve the coefficients $\sab$ are negligible for GPB.  (Note that a full analysis of the merits of dedicated light-bending tests for the SME was performed in Ref.\ \cite{TB11}.)

The leading aberration terms that arise in this calculation take the standard form.
Specifically we can write the accumulated change in the gyro spin due to aberration as
\begin{equation}
(\delta {\hat S}_{\text{\tiny NS}})_{\text{\tiny aberration}} = - \ens \cdot \vec v + \tfrac 12 (\ens \cdot \vec v) (\egs \cdot \vec v),
\label{aberration}
\end{equation}
with a similar equation holding for the WE direction.  To the necessary order, the velocity $\vec v$ can written in the SCF as $\vec v = \vec V_\oplus + \vec v_{\rm s}$, where $\vec V_\oplus$ is the velocity of the Earth's orbit around the Sun and $\vec v_{\rm s}$ is the velocity of the
gyroscope around the Earth.  For most of these aberration terms, the possible effects from the coefficients $\sab$ would arise through changes in the orbit via the perturbation terms in the equations of motion for the satellite (\ref{eom}).  If we average these terms over the time scale of the gyroscope orbit, we find that the possible effects due to the coefficients $\sab$ average to zero or are negligible, except for the term in (\ref{aberration}) that is linear in the Earth's velocity $\vec V_\oplus$.

The coefficients $\sab$ would also perturb the Earth's orbit.  For the time scale of the experiment, secular changes to the Earth's orbit are irrelevant and so the focus is on determining how oscillatory changes in the Earth's orbit might manifest in equation (\ref{aberration}).  The conventional annual variation of the leading aberration term is known and has a amplitude of about $20$~mas.  The oscillations of the Earth's orbit that would arise due to Lorentz violation in form of the $\sab$ coefficients, and could affect the velocity $\vec V_\oplus$ and hence equation (\ref{aberration}), would include both annual oscillations and twice annual oscillations \cite{BK06}.  Note that the actual GPB data collection time scale was just shy of one year.  It is therefore conceivable that GPB could be sensitive to a combination of $\sab$ coefficients via the oscillatory orbital effects on the Earth's velocity aberration terms.
On the other hand, the standard time dependence of the conventional aberration term was in fact used by the GPB experiment for calibration \cite{E11}.  It is therefore unclear whether constraints on such oscillatory effects could even be garnered from the GPB data.  This issue remains an open question for investigation but we remark that, regardless of this issue, our result constraining $\stt$ would remain unchanged since orbital effects are not sensitive to this coefficient at the post-newtonian order we consider in this work.

\section{Discussion} \label{sec:discussion}

We have used measurements of geodetic precession and frame-dragging by Gravity Probe~B to put the first direct experimental constraint on the time-time coefficient of Lorentz violation ($\stt$) in the Standard-Model Extension.
This coefficient controls Lorentz-violating effects associated with relativistic, post-Newtonian gravitational effects, 
and can also be measured by experiments involving light trajectories \cite{B09,TB11}.

Because the new constraint is linearly independent of existing limits on the spatial coefficients $\sjk$, it allows us to obtain individual upper bounds on $\stt,\sxx,\syy$ and $\szz$ for the first time.
These upper bounds strengthen slightly when a rescaling of Newton's gravitational constant in the theory is taken into account.  Our final upper bounds on the SME coefficients are given in Table~\ref{table:results}.

We have also considered orbital effects.
Our preliminary results suggest that these do not significantly affect the geodetic constraint from GPB, but can greatly strengthen the frame-dragging one, as given in Eq.~(\ref{gpbConstraintRevisedAgain}), so that it almost becomes competitive with laboratory limits on some of the spatial coefficients $\sjk$.
If these results are confirmed, then frame-dragging would appear to be an unexpectedly sensitive probe of Lorentz violation.
This possibility should be investigated further.
For instance, the inclusion of the Earth's quadrupole moment, both the conventional one from its rotation and a possible Lorentz-violating contribution due to spherical deformation, could affect the results in this work.
Also open for investigation is the issue of possible Lorentz-violating effects on the aberration terms arising from modifications to the Earth's orbit, 
as described in Sec.\ \ref{sec: add effects}.

Since the $\sab$ coefficients may have either sign, we note from Eqs.~(\ref{numericalLimits}) that the SME can accommodate precessions greater than, as well as less than those predicted by standard general relativity.
It shares this property with generalizations of Einstein's theory based on torsion \cite{M07}, but differs from others based on scalar fields or extra dimensions \cite{OEW13}, for which GR is a limiting case.

Future work can build on these results in various ways.
It would be of interest to study possible constraints from frame-dragging in other contexts, such as laser ranging to artificial satellites \cite{BK06,CP04} or signals from accretion disks around collapsed stars \cite{SV99}.
We have focused on the $\sab$ coefficients, but other sectors in the SME framework may also contribute, such as the matter-gravity coupling coefficients $a_\mu$ discussed in the literature \cite{KT09KT11}.
These coefficients are generically species-dependent, so extracting limits would likely require using test bodies of different composition \cite{H11H13}.

\begin{acknowledgments}
Thanks go to Ron Adler, Brett Altschul, Francis Everitt, Mac Keiser, Alan Kosteleck\'y, Barry Muhlfelder and Alex Silbergleit for comments.
Q.G.B. acknowledges Embry-Riddle Aeronautical University for support while this work was completed under Internal Award 13435.
J.M.O. and R.D.E. acknowledge the Department of Physics, Astronomy and Geosciences and the Fisher College of Science and Mathematics at Towson University respectively for travel support to present these results.
\end{acknowledgments}


\begin{thebibliography}{99}
\bibitem{Footnote2}We note that precessions in Ref.~\cite{BK06} are expressed in dimensionless form, so that the dimensional factor $gv_0$ does not appear in Eqs.~(154-160) of Ref.~\cite{BK06}, but is instead multiplied explicitly onto the equivalent of the relativistic drift equation~(\ref{GRdrift}) above. Here we prefer to express both precessions and relativistic drift rates in physical units of mas/yr. Also in Ref.~\cite{BK06}, the quantity denoted by $\alpha$ differs from $\ags$ by -$\pi$.
\bibitem{Footnote3}
There is a missing ``$-$'' sign in front of the expression for the unit vector $\hat{n}$ in terms of $\alpha$ on p.~32 of Ref.~\cite{BK06}.
\end{thebibliography}
\end{document}